\documentclass[prd,aps,showpacs,floats]{revtex4}
\usepackage{amsmath,amssymb}
\usepackage{graphicx}
\usepackage{epsfig}
\usepackage{array}
\usepackage{multirow}

\begin{document}

\title[Short Title]{New limits on  neutrino magnetic moment through non-vanishing 
13-mixing}
\author{M. M. Guzzo$^1$}\email{guzzo@ifi.unicamp.br} 
\author{P. C. de Holanda$^{1}$}\email{holanda@ifi.unicamp.br}
\author{O.  L. G. Peres$^{1,2}$}\email{orlando@ifi.unicamp.br} 
\affiliation{
  $^1$ Instituto de F\'\i sica Gleb Wataghin - UNICAMP, 
  13083-970 Campinas SP, Brazil \\
  $^2$ Abdus Salam International Centre for Theoretical Physics, ICTP, I-34010, Trieste, Italy
} 

\begin{abstract}

The relatively large value of neutrino mixing angle $\theta_{13}$
set by recent measurements allows us to use solar neutrinos to set a limit on 
neutrino magnetic moment involving second and third families, $\mu_{\mu\tau}$.
The existence 
of a random magnetic field in solar convective zone
can produce a significant anti-neutrino flux when a 
non-vanishing neutrino magnetic moment is assumed. 
%If the neutrino magnetic moment involves the first family, electron 
%anti-neutrinos are directly produced. However, for a neutrino magnetic moment 
%involving second and third families, 
Even if we consider a vanishing neutrino magnetic moment involving the 
first family, electron anti-neutrinos are indirectly 
produced through the mixing between first and third families and 
$\mu_{\mu\tau}\neq 0$. 
Using KamLAND limits on the solar flux of electron anti-neutrino, we set 
the limit $\mu_{\mu\tau}<0.5\times 10^{-11}\mu_B$ for a reasonable assumption on 
the behavior of solar magnetic fields.
This is the first time a limit on $\mu_{\mu\tau}$ is established in the 
literature directly from neutrino interaction with magnetic fields, 
and, interestingly enough, is comparable with the limits on neutrino magnetic 
moment involving the first family and with the ones coming from 
modifications on electroweak cross section.

\end{abstract}

\maketitle

%\baselineskip=20pt

%%%%%%%%%%%%%%%%%%%%%%%%%%%%%%%%%%%%%%%%%%%%%%%%%%%%%%%%%%%%%%%%%%
\section{Introduction}
%%%%%%%%%%%%%%%%%%%%%%%%%%%%%%%%%%%%%%%%%%%%%%%%%%%%%%%%%%%%%%%%%%%

In a recent paper~\cite{Guzzo:2005rr} we performed an analysis of how 
a non-vanishing neutrino transition magnetic moment involving second 
and third families, $\mu_{\mu\tau}$, 
could affect the flavour conversion of solar neutrinos. 
At that time we assumed a vanishing $\theta_{13}$, 
which allowed to produce a large flux 
of non-electronic anti-neutrinos, and our model was not limited 
by the absence of electron anti-neutrinos $\bar{\nu}_e$ 
in solar neutrino flux, as required by Kamland~\cite{Eguchi:2003gg}. 

However, in that paper it was argued that a non-vanishing 
$\theta_{13}$ would open a channel for the production of electron 
anti-neutrinos, and then a limit on $\mu_{\mu\tau}$ could be established 
from the  absence of a signal of $\bar{\nu}_e$ in solar neutrino flux. 
Since recent data indicates a 
relatively large value for this angle, we examine such limits in light of 
this new measurements.

\section{Conversion Probabilities}

To calculate the probability that a electron neutrino produced at the sun 
evolves into an electron anti-neutrino in the presence of transition 
magnetic moments, in principle we would have to work using 
a 6$\times$6 evolution matrix formalism, involving 
$\nu_a=(\nu_e,\nu_\mu,\nu_\tau,\bar{\nu}_e,\bar{\nu}_\mu,\bar{\nu}_\tau)^T$.
But the system can be simplified in specific cases. For instance, 
in~\cite{Guzzo:2005rr} we assumed a vanishing value for $\theta_{13}$, and 
rotating out the $23$-mixing with the definition $\nu'=U_{23}^{-1}\nu$, 
the system was decoupled into two 3$\times$3 
systems, which can be presented with a convenient reordering of eigenstates 
as:
\begin{eqnarray}
& 
i{\frac{\displaystyle{\ d }}{\displaystyle{\ dt}}}
\left(
\begin{array}{c}
\nu_e \\ \nu_\mu' \\ \bar{\nu}'_\tau \\ \bar\nu_e \\ \bar{\nu}'_\mu \\ \nu_\tau'  
\end{array}
\right) = 
& 
\left(
\begin{array}{cc}
    \left[\begin{array}{ccc}  & & \\& A & \\ & &\end{array}\right] &
    \begin{array}{ccc}  & & \\& 0 & \\ & &\end{array} \\
    \begin{array}{ccc}  & & \\& 0 & \\ & &\end{array} &
    \left[\begin{array}{ccc}  & & \\& \tilde{A} & \\ & &\end{array}\right] 
\end{array}
\right)
\left(
\begin{array}{c}
\nu_e \\ \nu_\mu' \\ \bar{\nu}'_\tau \\ \bar\nu_e \\ \bar{\nu}'_\mu \\ \nu_\tau'  
\end{array}
\right)\label{motion}
\end{eqnarray}
where
\begin{equation}
A=\left(
\begin{array}{ccc}
-\delta \cos2\theta_{12} +V_{CC}+V_{NC} & \delta \sin2\theta_{12}        & 0 \\ 
 \delta \sin2\theta_{12}           & \delta \cos2\theta_{12} + V_{NC} & \mu_{\mu\tau} B\\
 0              & \mu_{\mu\tau} B             & -\Delta - V_{NC}
\end{array}
\right)
\end{equation}
and $\tilde{A}$ is the same as $A$ with a change of sign on matter potentials. 
$\nu_\mu'$ and  $\nu_\tau'$ are linear combinations of weak
states as $\nu_\mu'=cos\,\theta_{23}\,\nu_\mu -sin\,\theta_{23}\,\nu_\tau$, 
$\nu_\tau'=sin\,\theta_{23}\,\nu_\mu +cos\,\theta_{23}\,\nu_\tau$, and with similar 
definitions to anti-neutrinos $\bar{\nu}$. Also, 
$\delta=\Delta m^2_{21}/4E$, $\Delta=\Delta m^2_{32}/4E$, and $V_{CC}$ and 
$V_{NC}$ are the charged current and neutral current interaction potentials 
with matter. 
Since all neutrinos in the sun are produced 
as electron neutrinos and the two systems are completely decoupled, no 
$\bar{\nu}_e$ was produced.
For a regular magnetic field in the convective zone 
of the order of $100$ kG and for magnetic moments of the order of 
$10^{-11}$ $\mu_B$ we do not expect any transition to anti-neutrinos, since 
\begin{equation}
\mu_{\mu\tau}B\sim5.8\times 10^{-15}~{\rm eV} << \Delta|_{E=10~{\rm MeV}}\sim 10^{-10}~{\rm eV}~.
\end{equation}
However, 
random fluctuations of magnetic fields in convective zone are 
expected and promote the 
population of anti-neutrino states families~\cite{Guzzo:2005rr}. 
This is implemented through symmetric
entries in Liouville equation, which induces decoherence, raising the $\nu\rightarrow\bar\nu$ conversion probability.

Nevertheless, 
when a non-vanishing value of $\theta_{13}$ is assumed, we can not decouple 
the system, and have to solve the full 6$\times$6 evolution equation. 
Rotating out both the mixing angles $\theta_{13}$ and $\theta_{23}$, 
we would have the following evolution matrix in the basis 
$\nu'=U_{13}^{-1}U_{23}^{-1}\nu$ conveniently rearranged as in Eq.~(\ref{motion}):

\begin{equation}
i{\frac{\displaystyle{\ d }}{\displaystyle{\ dt}}}\nu'=
\left(\begin{array}{cc}
\left[A\right]&\left[\tilde{B}\right] \\ \left[B\right] & \left[\tilde{A}\right]
\end{array}\right)\nu'
\end{equation}
with
\begin{equation}
A=
\left(\begin{array}{ccc}
-\delta \cos2\theta_{12}+c_{13}^2V_{CC}+V_{NC}& \delta \sin2\theta_{12} &  0 \\
\delta \sin2\theta_{12} & +\delta \cos2\theta_{12}+V_{NC}& \mu_{\mu\tau}Bc_{13}  \\
0 & \mu_{\mu\tau}Bc_{13}  & -\Delta - s^2_{13}V_{CC}-V_{NC}\\
\end{array}
\right)
\end{equation}
and
\begin{equation}
B=
\left(\begin{array}{ccc}
0 & -\mu_{\mu\tau}Bs_{13} & -s_{13}c_{13}V_{CC}  \\
- \mu_{\mu\tau}Bs_{13} & 0 & 0 \\
s_{13}c_{13}V_{CC} & 0 & 0
\end{array}
\right)~.
\end{equation}
where $s_{13}=\sin\theta_{13}$ and $c_{13}=\cos\theta_{13}$. 
$\tilde{A}$ ($\tilde{B}$) equals $A$ ($B$) by changing sign in matter potential 
terms. We will assume that the magnetic field is composed by a regular  part and a random part. Again, for a regular magnetic field we do not expect significant 
production 
of anti-neutrinos. However, assuming a random component of the magnetic field, 
anti-neutrinos can be produced through different channels.

To include the random magnetic fields with the same procedure, we should use 
the density  matrix formalism. From the $6\times 6$ Hamiltonian matrix, we get in the matrix density formalism a  35$\times$35 evolution system. 
Due to the complications of this procedure
we will present later the full analysis of the system~\cite{future}, 
but we can get a 
good estimative  assuming for now that 
the antineutrino production process will not be very different than in 
the scenario with vanishing $\theta_{13}$. Our procedure then would be to 
calculate the anti-neutrino production by the same assumption of last paper, 
a vanishing $\theta_{13}$, and then calculate the amount of $\bar{\nu}_e$ which
is present in this anti-neutrino state in accordance with the measured value 
of $\theta_{13}$~\cite{An:2012eh,Ahn:2012nd,Abe:2011fz}.

\section{Results}

When we considered in~\cite{Guzzo:2005rr} that $\theta_{13}=0$, 
the $\bar{\nu}_\tau'$ in 
Eq.~(\ref{motion}) was identical to the mass eigenstate $\bar{\nu}_3$. As 
mentioned before, we will consider that the inclusion of a non-vanishing 
$\theta_{13}$ will not strongly change this production probability. However, 
the distribution of such anti-neutrino in the mass eigenstates is now:
\begin{equation}
\bar{\nu}_\tau'=-(s_{13}c_{12})\nu_1-(s_{13}s_{12})\nu_2+(c_{13})\nu_3~.
\end{equation}
Averaging out all terms involving oscillation between different mass scales 
when calculating the probabilities and assuming large values of the other 
mixing angles, we can write the electron anti-neutrino 
production as:
\begin{equation}
P(\nu_e\to \bar{\nu}_e)\sim\sin^2\theta_{13} P(\nu_e\to \bar{\nu}_\tau')|_{\theta_{13}=0}~.
\end{equation}

KamLAND~\cite{Eguchi:2003gg} 
sets the strongest limit in the electronic anti-neutrino flux from the 
sun, given by $\phi(\bar\nu_e)<3.7\times 10^2$ cm$^{-2}$s$^{-1}$. 
Writing in terms of a production probability and using solar 
model labeled GS98 in~\cite{Serenelli:2011py} where  
$\phi_{^8B}=5.6\times 10^{6}$ cm$^{-2}$s$^{-1}$, we obtain  
an upper limit of the electronic anti-neutrino production of 
$P<6.6\times 10^{-5}$. 
Considering all recent measurements of 
$\sin^2\theta_{13}$~\cite{An:2012eh,Ahn:2012nd,Abe:2011fz}, 
we will use the value of best fit point~\cite{GonzalezGarcia:2012sz} 
$\sin^2\theta_{13}=2.5\times 10^{-2}$ in our 
analysis. This translates into a limit on 
anti-neutrino production of:
\begin{equation}
P(\nu_e\to \bar{\nu}'_\tau)|_{\theta_{13}=0}<2.7\times 10^{-3}~.
\label{eq:problimit}
\end{equation}

To calculate the anti-neutrino production probability we follow 
the procedure presented at~\cite{Guzzo:2005rr}. The probability is a function of 
the parameter 
\begin{equation}
k=<(\mu B_\perp)^2 > L_0
\end{equation}
where $L_0$ is a length scale related to the spatial coherence of the magnetic 
fluctuations. Rewriting $k$ in convenient units, we have:
\begin{equation}
k=3.4\times 10^{-17}\left[\frac{\mu}{10^{-11}\mu_B}\right]^2
\left[\frac{B}{100\,{\rm kG}}\right]\left[\frac{L_0}{200\,{\rm km}}\right]\,
{\rm eV}~.
\label{eq:k}
\end{equation}

We solved the evolution equation numerically assuming a vanishing $\theta_{13}$, as in~\cite{Guzzo:2005rr}. In Fig. 1  we present the 
conversion probability of anti-neutrinos 
if we assume a  vanishing $\theta_{13}$, together with the limits on 
this probability that can be inferred from KamLAND data and the measured 
values of $\theta_{13}$, as presented in Eq.~(\ref{eq:problimit}). The parameter 
region where the anti-neutrino production probability is larger than the 
KamLAND limit is excluded.

\begin{figure}[ht]
\begin{center}
\vspace{1cm}
\includegraphics[width=10cm]{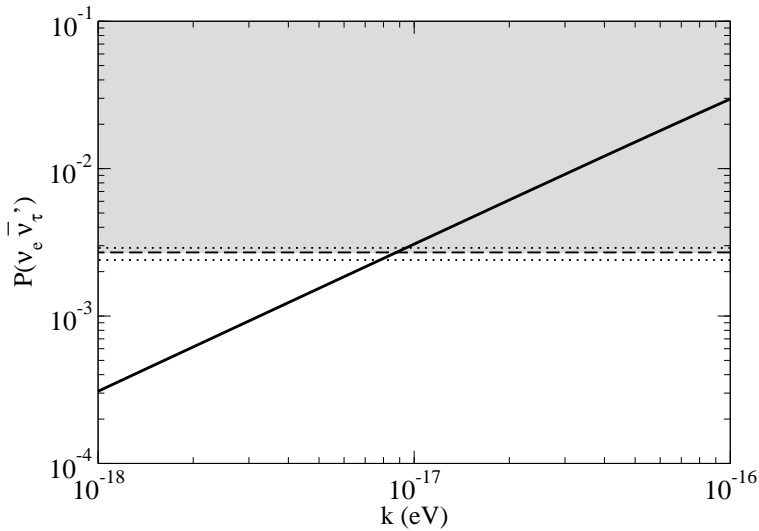}
\caption{We present here the comparison between KamLAND limits on 
anti-neutrino production with the predictions of such production in our model. 
The solid line corresponds to the anti-neutrino production 
probability in the Sun, in the assumption of a vanishing $\theta_{13}$. 
The dashed line corresponds to the limits set by KamLAND on solar electronic 
anti-neutrino flux, converted to a limit on non-electronic anti-neutrino 
probability conversion, using $\sin^2\theta_{13}=0.025\pm0.0023$ 
(1$\sigma$)~\cite{GonzalezGarcia:2012sz}.
The dotted lines are obtained with the 1$\sigma$ limits for this angle.}
\end{center}
\label{faca}
\end{figure}

From Fig. 1 we can extract a limit on $k$:
\begin{equation}
k<8\times10^{-18}\,{\rm eV}~,
\end{equation}
which leads, from Eq.~\ref{eq:k} to 
the following limit on magnetic field parameters:
\begin{equation}
\left[\frac{\mu}{10^{-11}\mu_B}\right]^2
\left[\frac{B}{100\,{\rm kG}}\right]\left[\frac{L_0}{200\,{\rm km}}\right]\,
{\rm eV}<0.24
\end{equation}

For a reasonable assumption on magnetic field profile, {\it i.e.} a 
$100$ kG regular magnetic field with random fluctuations proportional to the 
regular one, and a $200$ km coherent length scale for such fluctuations, 
we translate this limit to:
\begin{equation}
\mu_{\mu\tau}<0.5\times 10^{-11}\mu_B~.
\end{equation}

Such a limit can be compared with the ones coming from modifications on 
neutrino electroweak cross 
section~\cite{Grimus:2002vb,Liu:2004ny,Montanino:2008hu,Arpesella:2008mt}, 
which applies for a combination on all neutrino magnetic moment elements. 
Although our limit is more stringent then the one reported for instance 
in~\cite{Arpesella:2008mt}, it depends on both the solar magnetic field 
profile and the characteristics of its random fluctuations.

\section{Conclusions}

In this work we set a limit on the neutrino transition magnetic moment 
involving the second and third families using solar neutrino data and assuming 
a specific profile for the solar magnetic field. 
For a vanishing mixing angle 
$\theta_{13}$ we could only set loose bounds on such magnetic moment due to the 
electron anti-neutrino flavour decoupling on neutrino evolution equation. 
Now with a reasonable high measured value for such angle, a stringent limit was 
established, for the first time from the direct interaction of neutrinos with 
magnetic fields, at the same order of magnitude of the limits 
involving the first neutrino family and the limits coming from modifications on 
electroweak cross section.

Some approximations were made in the calculations that allowed us to use 
a previous study to calculate the anti-neutrino appearance probability. We 
plan to critically evaluate such approximations in a future work, but we expect 
that the limits established here are conservative, and a more detailed 
analysis could open other channels of anti-neutrino production, improving 
the limit presented here.

\section*{Acknowledgments}
The authors thank FAPESP and CNPq for several financial supports.

%%%%%%%%%%%%%references%%%%%%%%%%%%%%%%%%%%%%%%%%%%%%%%%%%%%%%%%%
%\bibliography{bibliography}{}
%\bibliographystyle{plain}

\end{document}